# A Comparative Study on Routing Selection Algorithms for Dynamic Planning of EONs over C+L Bands

F. Arpanaei[(1),(2)], J. M. Rivas-Moscoso[(2)], M. Ranjbar Zefreh[(3)], J. A. Hernández[(1)], J. P. Fernández-Palacios[(4)], D. Larrabeiti[(1)]

*(1) Department of Telematic Engineering, Universidad Carlos III de Madrid, 28911, Leganes, Madrid, Spain, farhad.arpanaei@uc3m.es*
*(2) Telefónica Research and Development, Ronda de la Comunicación, S/N, 28050 Madrid, Spain,*
*(3) CISCO Systems S.R.L., Vimercate (MB), Italy,*

**Abstract:** The performance of three routing selection algorithms is compared in terms of bandwidth blocking probability, quality of transmission, and run time in EONs over the C+L band. The min-max frequency algorithm shows the best performance on all metrics. © 2023 The Author(s).

## 1. Introduction

A major challenge in the design of dynamic elastic optical networks (EON) over C+L band is to consider the effect of inter-channel stimulated Raman scattering (ISRS), which causes power profile variations with frequency and distance, resulting in power transfer from C-band to L-band channels. On-line provisioning in dynamic networks require general, accurate, and sufficiently fast algorithms, in particular for channels with different modulation format levels (MFLs) and, therefore, varying maximum reach distances (MRDs). However, unlike for static C+L-band network planning, not much literature on the planning of dynamic C+L-band EONs is available. In [1], band prioritization schemes for spectrum allocation based on the generalized signal-to-noise ratio (GSNR) merit were investigated by using the generalized Gaussian noise (GGN) model, but the study did not take into account the effect of the MFL correction term. In [2] and [3], networking studies were proposed based on a four-wave mixing (FWM) closed-form formula accounting for multi-MFL/single Baud rate (BR) and single MFL/different BRs, respectively, but the effect of ISRS on the non-linear interference (NLI) Kerr effect was not considered. In all cases, the path selection algorithm was based on the K-shortest path with available continuous, contiguous spectrum. In this paper, we propose a generic and fast routing, modulation format, and spectrum assignment (RMSA) tool for C+L-band EONs using the ISRS closed-form formula in [4] to calculate the optimal power and reach for different MFLs, whereby we compare 3 RMSA algorithms based on the calculation of the *K*-shortest path that minimize maximum frequency (MaxF) and maximum number of hops (MaxHop), and maximize the minimum GSNR (MinGSNR) merit.

## 2. System Model and Proposed Routing Selection Algorithms

A C+L-band EON can be modeled as a graph with vertices (nodes) and edges (links). The nodes are equipped with C+L-band transponders and reconfigurable add-drop multiplexers (ROADM). The links are made up of a sequence of spans of average length *L*, with in-line C+L-band erbium-doped fiber amplifiers (EDFA) equipped with gain-flattening equalizers between them. The end-to-end GSNR of a channel with launch power $P_\text{ch}$ can be estimated as $GSNR^{-1} \approx \sum_{l=1}^{N_\text{L}} \sum_{s=1}^{N_\text{s}} [(P_\text{ch} - P_\text{NLI}^{s,l})/(P_\text{ASE}^{s,l} + P_\text{NLI}^{s,l})]^{-1} + (SNR_\text{TRx})^{-1}$, where $N_\text{L}$, $N_\text{s}$, $P_\text{ASE}^{s,l}$, $P_\text{NLI}^{s,l}$, and $SNR_\text{TRx}$ are the number of links (*l*), the number of spans (*s*) on those links, the noise power of the EDFAs, the noise power due to the ISRS and NLI Kerr effects, and the back-to-back implementation penalty, respectively. $P_\text{ASE}^{s,l}$ and $P_\text{NLI}^{s,l}$ are obtained from (1) and (2) in [4], and we apply (2) of [5] to estimate the optimum launch power for the network (assuming a uniform power profile) and the MRD for each MFL that guarantee the GSNR threshold is met across the entire C+L bands. Our proposed RMSA algorithm comprises eight steps: ***step1***- *K* shortest paths between all source-destination (*src, des*) pairs (in the following called "connections") are pre-calculated, including attributes such as $N_\text{L}$, $N_\text{s}$ of each link, and the list of links; ***step2***- the nodal-degree probability distribution function (PDF) of the network for each connection is estimated in advance as $PDF = D_{src}D_{des}/(\sum_{i=1}^{N-1}\sum_{j=i+1}^{N} D_i D_j)$, where $D_i$ and $D_j$ are the nodal degrees of nodes *i* and *j*. After the pre-processing steps, in ***step3***, for each arriving request, a connection is randomly selected based on its nodal-degree PDF and, in ***step4***, the request throughput is randomly generated based on a uniform distribution with values spanning [100, 600] Gb/s, with step 100 Gb/s; next, in ***step5***, the MFL (*m*) with throughput $R_m$ is estimated according to a lookup table obtained from (2) in [5], and, in ***step6***, the number of required channels for the request *r* with throughput $R_r$ over connection *(src,des)* is calculated from $N_r = \lceil R_r/R_m \rceil \times \lceil R_{symb}/12.5 \rceil$, where $R_{symb}$ is the selected symbol rate in GBaud; ***step7***- the spectrum assignment is performed based on first-fit algorithm and spectrum continuity constraints for all pre-calculated K-shortest paths, and the MaxF, MaxHop and MinGSNR of the assigned channels for all feasible paths are stored; ***step8***- the best path is selected based on the chosen path priority, including (1) minimum MaxF (MinMaxF), (2) minimum MaxHop (MinMaxHop), or (3) maximum MinGSNR (MaxMinGSNR).

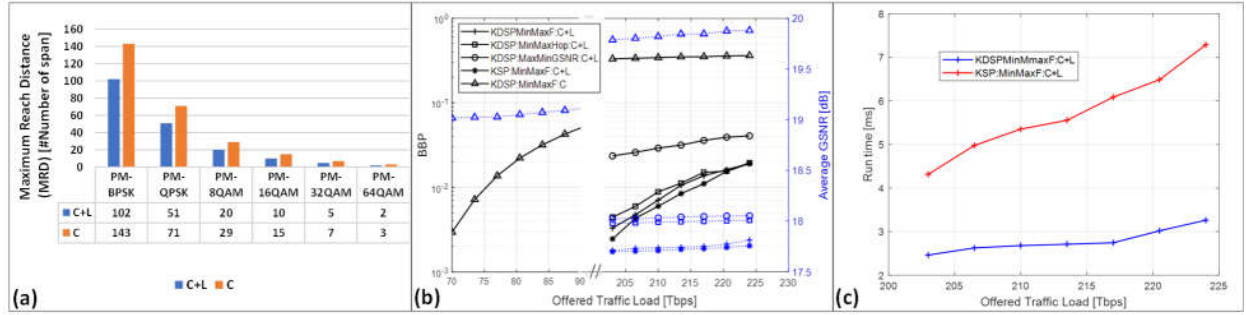

Fig. 1. (a) MRDs for C and C+L scenarios, (b) bandwidth blocking probability (BBP) and average GSNR, (c) run time of KSP and KDSP. KDSP: K-disjoint shortest path, KSP: K-shortest path.

## 3. Simulations, Results, and Conclusion

We use the above system model to calculate the optimal channel power and carry out a network-level analysis over a C+L-band EON based on the national network of Telefónica Spain, composed of 14 nodes and 22 links, with an average span length of 70 km (ITU-T G.652.D fiber), and benchmark results against a C-band-based EON following the LOGON strategy [6] for the central channel (worst case). We consider coherent transponders operating at a fixed BR of 64 GBaud with $SNR_{TRx}$ = 36 dB and spectral bandwidth of 75 GHz; 2-dB aging SNR margin, EDFA noise figure of 4.5 and 6 dB in C and L band, respectively; bit error rate (BER) threshold of $1 \times 10^{-3}$, suitable for a 28% forward error correction (FEC) overhead; and, like in [1], a guard band between the C and L bands of 500 GHz. With these values, the GSNR thresholds for MFL PM-BPSK, PM-QPSK, and PM- 8/16/32/64QAM (corresponding to bit rates ranging from 100 to 600 Gb/s) are 6.79, 9.81, 13.71, 16.54, 19.58, and 22.54 dB, respectively. We solve the optimization algorithm presented in Eq. (2) of [5] to find the MRD lookup table for each MFC accounting for ISRS and Kerr NLI (Fig. 1.a) and the optimum launch power for all listed MCFs, which amounts to 0.1 and -0.15 dBm, respectively, for C and C+L band, assuming fully loaded links (53 and 91 channels in C and L bands, respectively). The arrival time (AT) and holding time (HT) of each request follows a Poisson distribution with an average arrival rate of $\mu_{AT}$ requests per unit of time and a negative exponential distribution with an average of $1/\mu_{HT}$, respectively, and no restriction is imposed on the number of transceivers per node to serve those requests. The offered traffic load is then $\mu_{AT}/\mu_{HT}$ (measured in normalized traffic units), and, for each load, the bandwidth blocking probability (BBP) is calculated with 95% confidence. Candidate connections are generated for the K-shortest paths (KSP) and disjoint paths (KDSP), with $K$ = 5, and the best connection is selected among them, while ASE-noise loading is considered for unused channels so that the GSNR thresholds estimations above are valid. Despite the presence of ISRS, the network throughput suffers no penalty, as depicted in Fig.1.b, when transitioning from C- to C+L band solutions (for BBP=$10^{-2}$, the throughput increases proportionally with the spectrum). The MinMaxF metric outperforms other metrics in terms of BBP and GSNR for both KSP and KDSP, with KSP showing a slightly better outcome. However, the run time for KSP is significantly higher than for KDSP, especially for increasing offered loads, as shown in Fig.1.c. To obtain a more comprehensive understanding of the performance of the proposed RMSA algorithms, topologies with varying sizes must be investigated and compared.

**Acknowledgement.** Farhad Arpanaei acknowledges support from the CONEX Plus programme funded by Universidad Carlos III de Madrid and the European Union's Horizon 2020 research and innovation programme under the Marie Skłodowska-Curie grant agreement No. 801538. The authors would like to acknowledge the support of EU-funded B5G-OPEN and ALLEGRO projects (grant No. 101016663 and 101092766).

## 4. References

[1] N. Sambo *et al.*, "Provisioning in Multi-Band Optical Networks," *J. Light. Technol.*, 38 (9), 2598-2605, 2020.
[2] D. Uzunidis, *et al.*, "Strategies for Upgrading an Operator's Backbone Network Beyond the C-Band: Towards Multi-Band Optical Networks," *IEEE Photon. J.*, 13 (2), 1-18, 2021.
[3] D. Uzunidis *et al.,* "Connectivity Challenges in E, S, C and L Optical Multi-Band Systems," ECOC 2021.
[4] D. Semrau *et al.,* "Modulation format dependent closed-form formula for estimating nonlinear interference in s+c+l band system," ECOC 2019.
[5] F. Arpanaei *et al.*, "Launch Power Optimization for Dynamic Elastic Optical Networks over C+L Bands," ONDM 2023, doi: https://doi.org/10.5281/zenodo.7733899.
[6] P. Poggiolini *et al.*, "Recent Advances in the Modeling of the Impact of Nonlinear Fiber Propagation Effects on Uncompensated Coherent Transmission Systems," *J. Light. Technol.*, 35(3), 458-480, 2017.